\def\year{2016}\relax
\documentclass[letterpaper]{article}
\usepackage{aaai16}
\usepackage{times}
\usepackage{helvet}
\usepackage{courier}

\usepackage{subfigure}
\usepackage{graphicx,url}
\usepackage{fancyhdr}
\usepackage{rotating}
\usepackage{subfigure}
\usepackage{amssymb}
\usepackage{amsmath}
\usepackage{multirow}
\usepackage{multicol}
\usepackage{indentfirst}
\usepackage{float}
\usepackage{epsfig}
\usepackage{algorithm,algorithmic}
\usepackage{subfigure}
\usepackage{epstopdf}
\usepackage{epsfig}
\usepackage{amsmath}
\usepackage{comment}
\usepackage{color}

\usepackage[labelfont=bf]{caption}

\usepackage{color}

\usepackage{booktabs} 

\begin{document}
%
\title{Emotions, Demographics and Sociability in Twitter Interactions} 
\author{
Kristina Lerman\\
USC Information Sciences Institute\\
Marina del Rey, CA 90292 \\
\texttt{lerman@isi.edu}
\And
Megha Arora \\
IIIT-Delhi\\
New Delhi, India - 110020 \\
\texttt{megha12059@iiitd.ac.in}
\And
Luciano Gallegos\\
USC Information Sciences Institute\\
Marina del Rey, CA 90292 \\
\texttt{luciano.gallegos@gmail.com}
\AND
Ponnurangam Kumaraguru\\
IIIT-Delhi\\
New Delhi, India - 110020 \\
\texttt{pk@iiitd.ac.in}
\And
David Garcia\\
ETH-Zurich\\
Zurich, Switzerland\\
\texttt{dgarcia@ethz.ch}
}
\maketitle
\begin{abstract}
The social connections people form online affect the quality of information they receive and their online experience. Although a host of socioeconomic and cognitive factors were implicated in the formation of offline social ties, few of them have been empirically validated, particularly in an online setting. In this study, we analyze a large corpus of geo-referenced messages, or tweets, posted by social media users from a major US metropolitan area. We linked these tweets to US Census data through their locations. This allowed us to measure emotions expressed in the tweets posted from an area, the structure of social connections, and also use that area's socioeconomic characteristics in analysis.  
We find that at an aggregate level, places where social media users engage more deeply with less diverse social contacts are those where they express more negative emotions, like sadness and anger. Demographics also has an impact: these places have residents with lower household income and education levels. Conversely, places where people engage less frequently but with diverse contacts have happier, more positive messages posted from them and also have better educated, younger, more affluent residents. Results suggest that cognitive factors and offline characteristics affect the quality of online interactions. Our work highlights the value of linking social media data to traditional data sources, such as US Census, to drive novel analysis of online behavior.

\end{abstract}




\section{Introduction}
Humans have evolved large brains, in part to handle the cognitive demands of social relationships~\cite{Dunbar03}. The social structures resulting from these relationships confer numerous fitness advantages.  Scholars distinguish between two types of social relationships: those representing strong and weak ties. Strong ties are characterized by high frequency of interaction and emotional intimacy that can be found in relationships between family members or close friends. People connected by strong ties share mutual friends~\cite{granovetter1973}, forming cohesive social bonds that are essential for providing emotional and material support~\cite{Putnam2000bowling,Rime2009} and creating resilient communities~\cite{sampson1997neighborhoods}.
In contrast, weak ties represent more casual social relationships, characterized by less frequent, less intense interactions, such as those occurring between acquaintances.  By bridging otherwise unconnected communities, weak ties  expose individuals to novel and diverse information that leads to new job prospects~\cite{Granovetter83} and career opportunities~\cite{Burt95,Burt04}.
Online social relationships provide similar benefits to those of the offline relationships, including emotional support and exposure to novel and diverse information~\cite{Aral11,Bakshy2012therole,Kang15icwsm}.

How and why do people form different social ties, whether online or offline? Of the few studies that addressed this question, Shea et al.~\cite{Shea15} examined the relationship between emotions and cognitive social structures, i.e., the mental representations individuals form of their social contacts~\cite{krackhardt1987cognitive}.
In a laboratory study, they demonstrated that subjects experiencing positive affect, e.g., emotions such as happiness, were able to recall a larger number of more diverse and sparsely connected social contacts than those experiencing negative affect, e.g., sadness. In other words, they found that positive affect was more closely associated with weak ties  and negative affect with strong ties in cognitive social structures. This is consistent with findings that negative emotional experiences are shared more frequently through strong ties~\cite{Rime2009}, not only to seek support but also as a means of strengthening the tie~\cite{Niedenthal2012}.
In addition to psychological factors, social structures also depend on the participants' socioeconomic and demographic characteristics. A study, which reconstructed a national-scale social network from the phone records of people living in the United Kingdom, found that people living in more prosperous regions formed more diverse social networks, linking them to others living in distinct communities~\cite{Eagle:Network2010}. On the other hand, people living in less prosperous communities formed less diverse, more cohesive social structures.

The present paper examines how psychological and demographic factors affect the structure of online social interactions.
We restrict our attention to interactions on the Twitter microblogging platform. To study these interactions, we collected a large body of geo-referenced text messages, known as tweets, from a large US metropolitan area.  Further, we linked these tweets to US Census tracts through their locations. Census \emph{tracts} are small regions, on a scale of city blocks, that are relatively homogeneous with respect to population characteristics, economic status, and living conditions.
Some of the tweets also contained explicit references to other users through the `@' mention convention, which has been widely adopted on Twitter for conversations. We used mentions to measure the strength of social ties of people tweeting from each tract.
Using these data we studied (at tract level) the relationship between social ties, the socioeoconomic characteristics of the tract, and the emotions expressed by people tweeting from that tract. In addition, people tweeting from one tract often tweeted from other tracts. Since geography is a strong organizing principle, for both offline~\cite{Milgram69,Barthelemy20111} and online~\cite{Quercia:Tracking2012,LibenNowell,Backstrom} social relationships, we measured the spatial diversity of social relationships, and studied its dependence on socioeconomic, demographic, and psychological factors.

Our work complements previous studies of offline social networks and demonstrates a connection between the structure of online interactions in urban places and their socioeconomic characteristics. More importantly, it links the structure of online interactions to positive affect. People who express happier emotions interact with a more diverse set social contacts, which puts them in a position to access, and potentially take advantage of, novel information. As our social interactions increasingly move online, understanding, and being able to unobtrusively monitor, online social structures at a macroscopic level is important to ensuring equal access to the benefits of social relationships.

In the rest of the paper, we first describe data collection and methods used to measure emotion and social structure. Then, we present results of a statistical study of social ties and their relationships to emotions and demographic factors. The related works are addressed after this.
Although many important caveats exist about generalizing results of the study, especially to offline social interactions, our work highlights the value of linking social media data to traditional data sources, such as US Census, to drive novel analysis of online behavior and online social structures.

\section{Related Work} \label{sec:related}   

Eagle et al.~\cite{Eagle:Network2010} explored the link between socioeconomic factors and network structure using anonymized phone call records to reconstruct the national-level network of people living in the UK. Measures of socioeconomic development were constructed from the UK government's Index of Multiple Deprivation (IMD), a composite measure of prosperity based on income, employment, education, health, crime, housing of different regions within the country. They found that people living in more prosperous regions formed more diverse social networks, linking them to others living in distinct communities. On the other hand, people living in less prosperous communities formed less diverse, more cohesive social structures.


Quercia et al.~\cite{Quercia:Tracking2012} found that sentiment expressed in tweets posted around 78 census areas of London correlated highly with community socioeconomic well being, as measured by the Index of Multiple Deprivation (i.e., qualitative study of deprived areas in the UK local councils). In another study~\cite{Quercia2012social} they found that happy places tend to interact with other happy places, although other indicators such as demographic data and human mobility were not used in their research~\cite{Cheng:Footprints2011}.

Other researcher used demographic factors and associated them to sentiment analysis to measure happiness in different places.
For instance, Mitchell et al.~\cite{Mitchell:Happiness2013} generated taxonomies of US states and cities based on their similarities in word use and estimates the happiness levels of these states and cities.
Then, the authors correlated highly-resolved demographic characteristics with happiness levels and connected word choice and message length with urban characteristics such as education levels and obesity rates, showing that social media may potentially be used to estimate real-time levels and changes in population-scale measures, such as obesity rates.

Psychological and cognitive states affect the types of social connections people form and their ability to recall them~\cite{Brashears13}. When people experience positive emotions, or affect, they broaden their cognitive scope, widening the array of thoughts and actions that come to mind~\cite{fredrickson2001role}. In contrast, experiencing negative emotions narrow attention to the basic actions necessary for survival.
Shea et al.~\cite{Shea15} tested these theories in a laboratory, examining the relationship between emotions and the structure of networks people were able to recall. They found that subjects experiencing positive affect were able to recall a larger number of more diverse and sparsely connected social contacts than those experiencing negative emotions. The study did not resolve the question of how many of the contacts people were able to recall that they proceeded to actively engage.

A number of innovative research works attempted to better understand human emotion and mobility. Some of these works focuses on geo-tagged location data extracted from Foursquare and Twitter.
Researchers reported~\cite{Cramer:Performing2011,Noulas:Foursquare2011} that Foursquare users usually check-in at venues they perceived as more interesting and express actions similar to other social media, such as Facebook and Twitter.
Foursquare check-ins are, in many cases, biased: while some users provide important feedback by checking-in at venues and share their engagement, others subvert the rules by deliberately creating unofficial duplicate and nonexistent venues~\cite{Duffy:Foursquare2011}.

\section{Methods} \label{sec:method}

\subsection{Data}
\label{sec:data}
\noindent
Los Angeles (LA) County is the most populous county in the United States, with almost 10 million residents. It is extremely diverse both demographically and economically, making it an attractive subject for research. We collected a large body of tweets from LA County over the course of 4 months, starting in July 2014.  Our data collection strategy was as follows. First, we used Twitter's location search API to collect tweets from an area that included Los Angeles County.  We then used Twitter4J API to collect all (timeline) tweets from users who tweeted from within this area during this time period. A portion of these tweets were geo-referenced, i.e.  they had geographic coordinates attached to them.
In all, we collected 6M geo-tagged tweets made by 340K distinct users.


We localized geo-tagged tweets to tracts from the 2012 US Census.\footnote{American Fact Finder (http://factfinder.census.gov/)} A tract is a geographic region that is defined for the purpose of taking a census of a population, containing about 4,000 residents on average, and is designed to be relatively homogeneous with respect to demographic characteristics of that population. We included only Los Angeles County tracts in the analysis. 
We used data from the US Census to obtain demographic and socioeconomic characteristics of a tract, including the mean household income, median age of residents, percentage of residents with a bachelor's degree or above, as well as racial and ethnic composition of the tract.

\subsection{Emotion Analysis}
\noindent

To measure emotions, we apply sentiment analysis~\cite{Pang2008}, i.e. methods
that  process text to quantify subjective states of the author of the text.
Two recent independent benchmark studies evaluate a wide variety of sentiment
analysis tools in various social media~\cite{Gonccalves2013} and
Twitter datasets~\cite{Abbasi2014}.  Across social media, one of the best
performing tools is SentiStrength~\cite{Thelwall2012}, which also was shown to
be the best unsupervised tool for tweets in various
contexts~\cite{Abbasi2014}.

SentiStrength quantifies emotions expressed in short informal text by
matching terms from a lexicon and applying intensifiers, negations,
misspellings, idioms, and emoticons.  We use the standard English version of
SentiStrength\footnote{http://sentistrength.wlv.ac.uk/} to each tweet in our
dataset, quantifying positive sentiment $P\in[+1,+5$] and negative sentiment
$N\in[-1,-5$], consistently with the Positive and Negative Affect Schedule
(PANAS)~\cite{Watson1988}.  SentiStrength has been shown to
perform very closely to human raters in validity tests~\cite{Thelwall2012} and
has  been applied to measure emotions in product reviews~\cite{Garcia2011},
online chatrooms~\cite{Garas2012},  Yahoo answers~\cite{Kucuktunc2012}, and
Youtube comments~\cite{Garcia2012}. In addition, SentiStrength allows our
approach to be applied in the future to other languages, like
Spanish~\cite{Alvarez2015}, and to include contextual
factors~\cite{Thelwall2013}, like sarcasm~\cite{Rajadesingan2015}.

Beyond positivity and negativity, meanings expressed through text can be
captured through the application  of the semantic
differential~\cite{Osgood1964}, a dimensional approach that quantifies
emotional meaning in terms of valence, arousal, and
dominance~\cite{Russell1977}. The dimension of \emph{valence} quantifies the
level of pleasure or evaluation expressed by a word,  \emph{arousal} measures
the level of activity induced by the emotions associated with a word, and
\emph{dominance} quantifies the level of subjective power or potency
experienced in relation to an emotional word.  Research in psychology suggests
that a multidimensional approach is necessary to capture the variance of
emotional experience~\cite{Fontaine2007}, motivating our three-dimensional
measurement beyond simple polarity approximations.
The state of the art in the quantification of these three dimensions is the
lexicon of Warriner, Kuperman, and Brysbaert (WKB)~\cite{Warriner2013}. The
WKB lexicon includes scores in the three dimensions for more than 13,000 English
lemmas.  We quantify these three dimensions in a tweet by first lemmatizing
the words in the tweet, to then match the lexicon and compute mean values of
the three dimensions as in \cite{Gonzalez2012}. The large size of this lexicon
allows us to match terms in in 82.39\% of the tweets  in our dataset, which we
aggregate to produce multidimensional measures of emotions.

\begin{figure*}[tbp]
\centering
\begin{tabular}{cc}  
  \includegraphics[width=0.85\columnwidth]{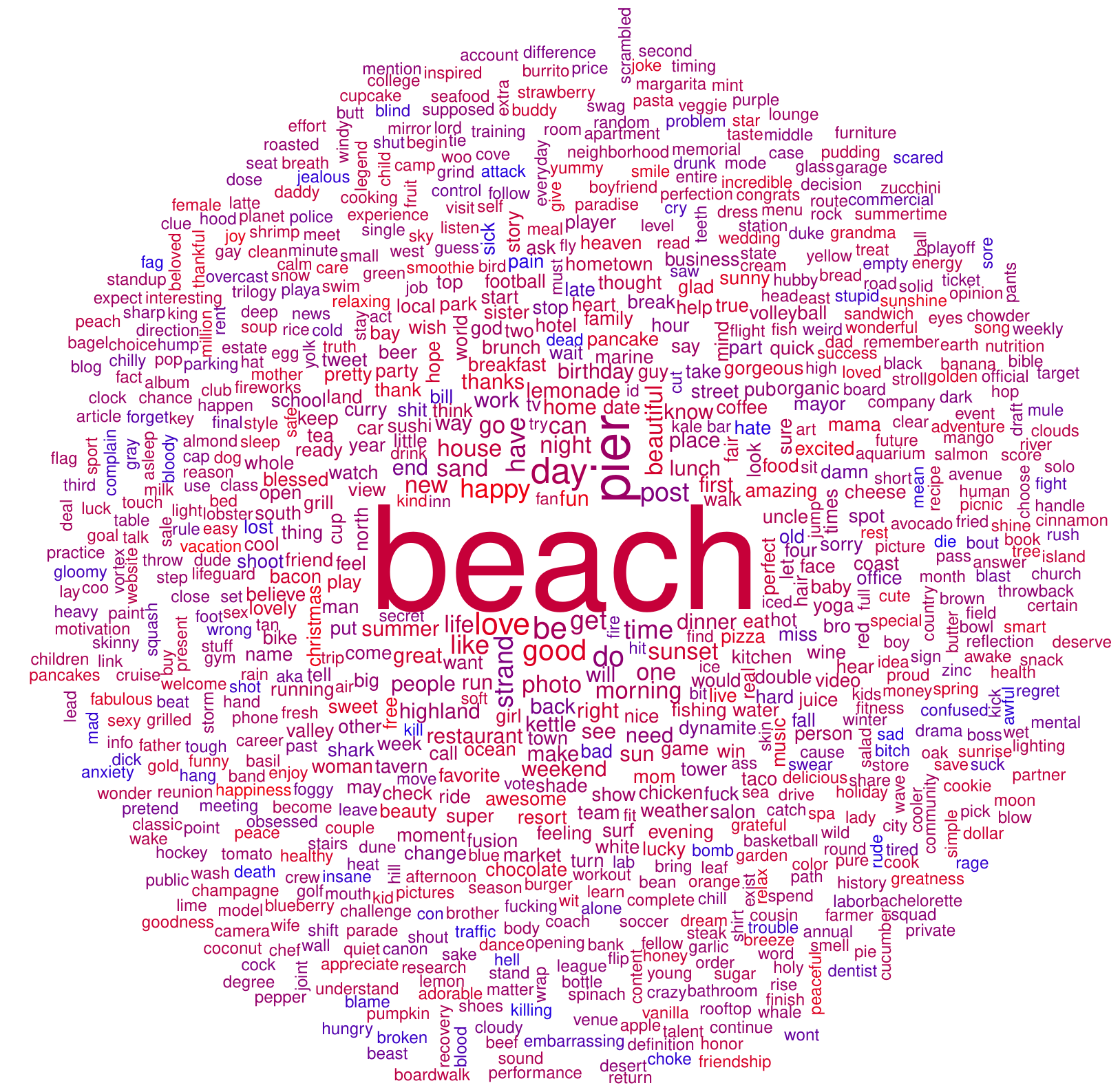}  
  &
  \includegraphics[width=0.9\columnwidth]{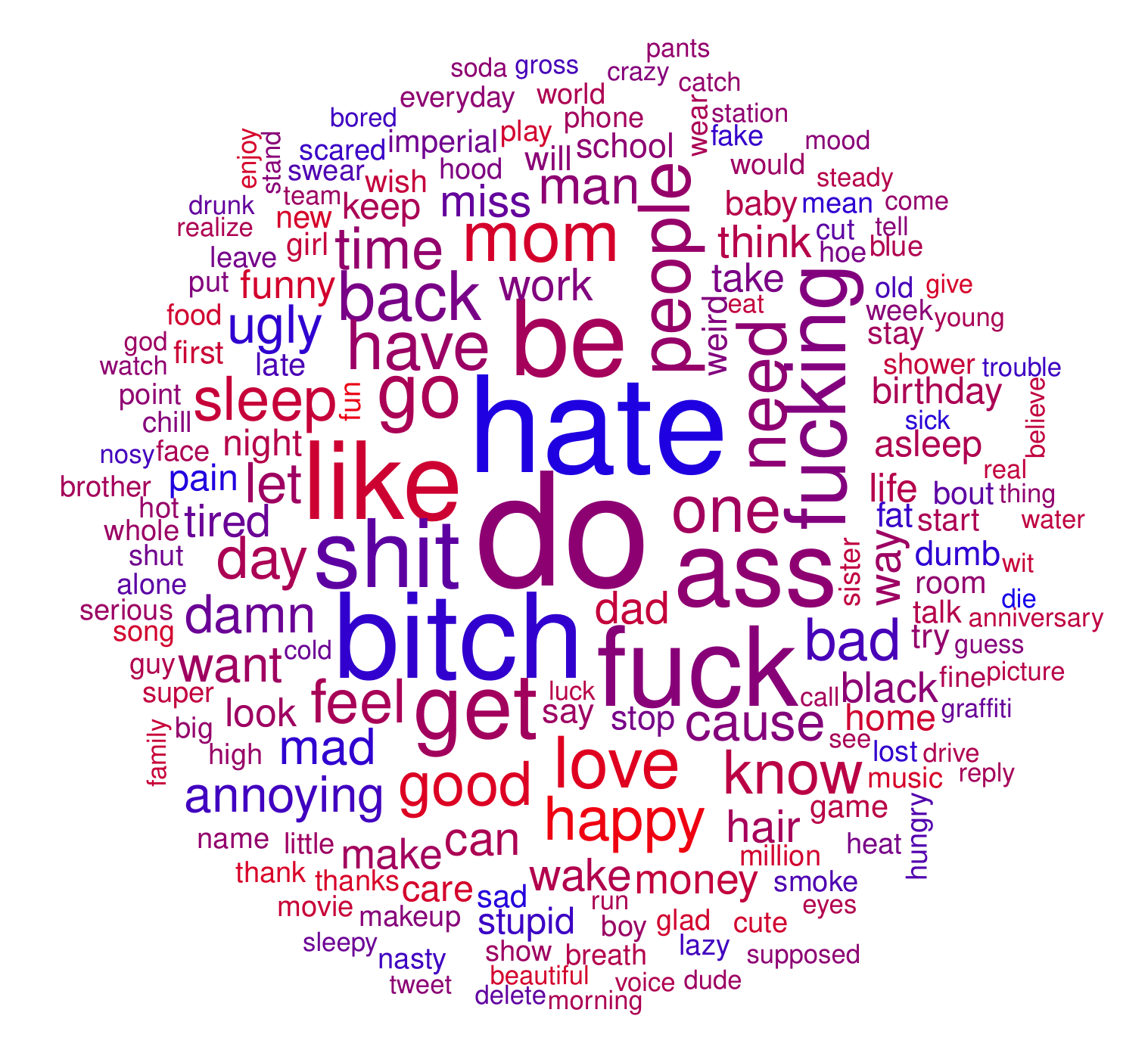} \\ 
 (a) & (b)
\end{tabular}
  \caption{Word cloud of tweets from two tracts with average valence of (a) 6.122 and (b) 5.418. Words are colored by their valence, with red corresponding to high valence words, and blue to low valence words.}\label{fig:wordle}
\end{figure*}

The Figure~\ref{fig:wordle} presents word clouds of tweets from a tract with one of the highest average valence and one from a tract with a lower average valence. The words themselves are colored by their valence, with red corresponding to high and blue to low valence words.
Despite seemingly small differences in average tract valence, the words depicted in the word clouds are remarkably different in the emotions they convey. The ``happy'' tract has words such as `beach', `love', `family', `beautiful', while the ``sad'' tract contains many profanities (though it also contains some happy words).



\subsection{Social Tie Analysis}
\label{sec:method-sna}
Twitter users address others using the `@' mention convention. We use the mentions as evidence of social ties, although sometimes users address public figures and celebrities also using this convention. We use mention frequency along a tie as a proxy of tie strength, drawing upon multiple studies that used frequency of interactions as a measure of tie strength~\cite{Granovetter83,onnela2007structure,Quercia:Tracking2012}.
In contrast to other measures, such as clustering coefficient, it does not require knowledge of full network structure (which we do not observe).

\begin{figure}[tbh]
  \centering
  \begin{tabular}{@{}c@{}c@{}}  
  \includegraphics[width=0.5\columnwidth]{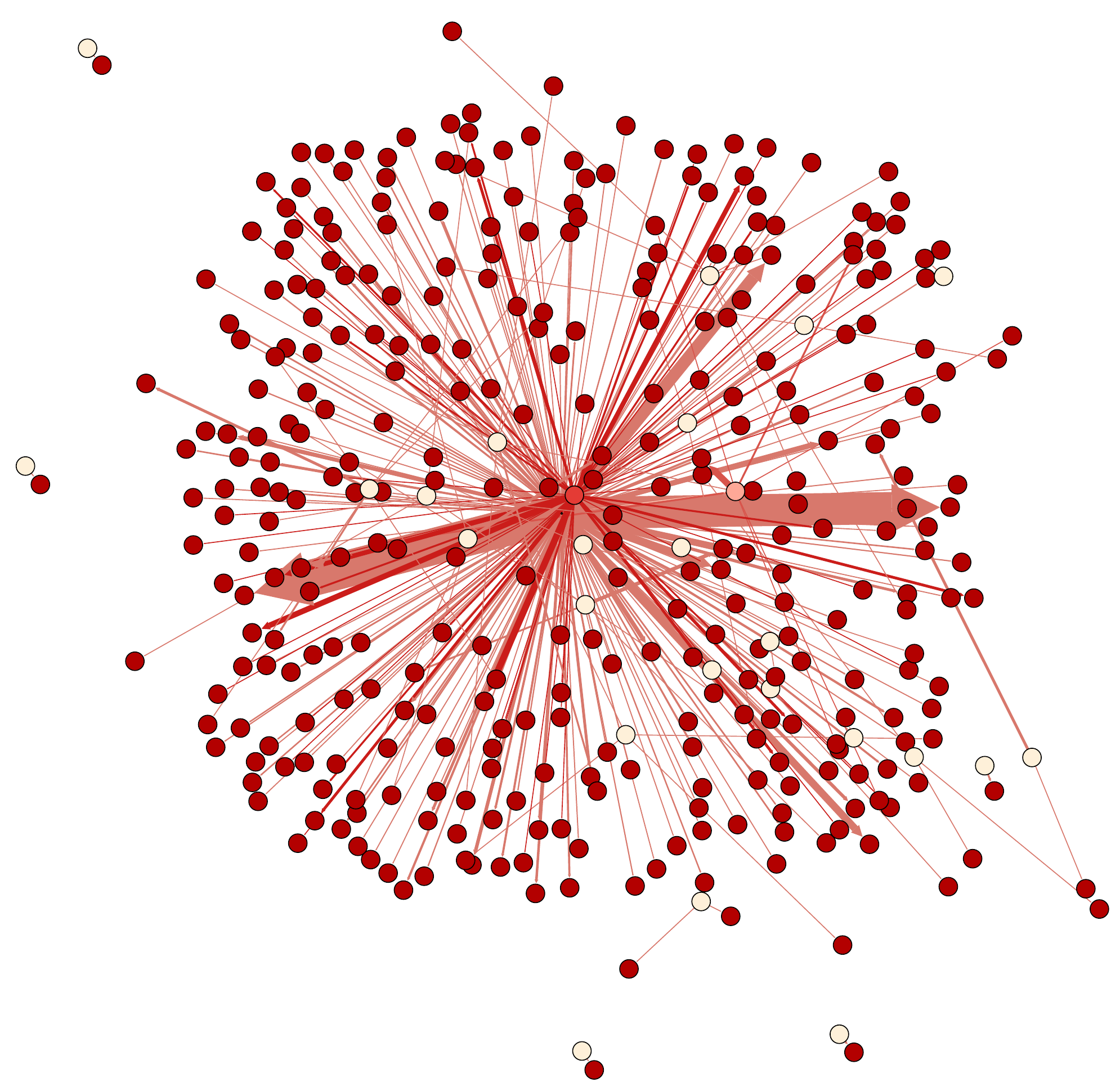}
  &
  \includegraphics[width=0.5\columnwidth]{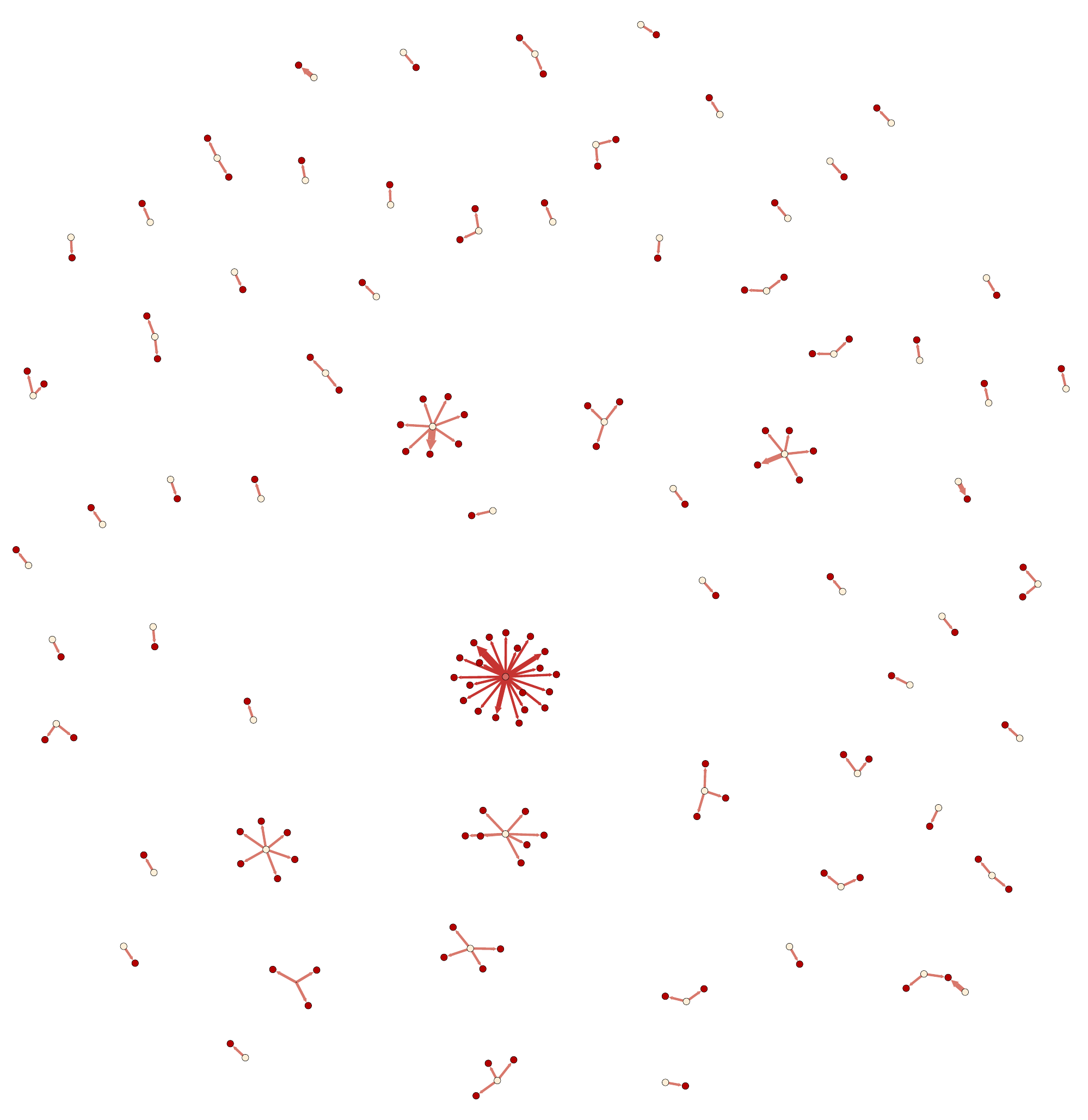} \\
  (a) & (b)
  \end{tabular}
  \caption{Mentions graphs of two different tracts showing (a) strong ties (average tie strength $S_i=7.33$) and (b) weak ties ($S_i=1.08$). Tweeting users are represented as white nodes, while mentioned users are red nodes. Users who tweet and are mentioned are pink in color. The width of an edge represents the number of mentions. }\label{fig:tie-strength-examples}
\end{figure}

\paragraph{Tie strength}
For each tract, we create a mention graph with users as nodes and an edge from user $A$ to user $B$ if $A$ mentions $B$ in her tweets.
Using this graph, the average social tie strength per tract is defined as
\begin{equation} \label{eq:st}
	S_{i} = \frac{\sum_{j=1}^{k_i}w_{j}}{k_{i}}
\end{equation}
\noindent
where $w_{j}$ is the weight of the $j^{th}$ edge (i.e., the number of times user $A$  mentioned user $B$), and $k_{i}$ is the total number of distinct users mentioned in tract $i$.

We do not have complete knowledge of network structure, since we only observe the tweets of users who geo-referenced their tweets, and not necessarily the tweets of mentioned users. However, even in the absence of complete information about interactions, average tie strength captures the amount of social cohesion and diversity~\cite{Gonccalves2011}.
Figure~\ref{fig:tie-strength-examples} illustrates mention graphs from two tracts with very different tie strength values. High tie strength (Fig.~\ref{fig:tie-strength-examples}(a)) is associated with a high degree of interaction and more clustering~\cite{granovetter1973}. In contrast, low tie strength is associated with a sparse, more diverse network with few interconnections (Fig~\ref{fig:tie-strength-examples}(b)).

\paragraph{Spatial diversity}
Geography and distance are important organizing principles of social interactions, both offline~\cite{Milgram69,Barthelemy20111} and online~\cite{Quercia:Tracking2012,LibenNowell,Backstrom}.
While most social interactions are short-range, long-distance interactions serve as evidence of social diversity~\cite{Eagle:Network2010}.
In this paper, we use the movement of people across tracts as evidence of the spatial diversity of their social structures. Following  Eagle et al.~\cite{Eagle:Network2010}, we measure spatial diversity of places from which people tweeting from a given tract also tweet from, using Shannon's Entropy ratio, as
\begin{equation} \label{eq:sd}
	D_{i} = \frac{-\sum_{j=1}^{n_i}p_{ij}log(p_{ij})}{\log n_i}
\end{equation}
\noindent
where $n_i$ is the number of tracts from which users who tweeted from tract $i$ also tweeted from, and $p_{ij}$ is the proportion of tweets posted by these users from tract $j$ such that
\begin{equation} \label{eq:sdp}
	p_{ij} = \frac{T_{ij}}{\sum_{j=1}^{n_i}T_{ij}}
\end{equation}
\noindent
where $T_{ij}$ is the number of tweets that have been posted in tract $j$ by the users who have tweeted from both tract $i$ and $j$.

Thus, spatial diversity is a ratio that compares the empirical entropy of data with its expected value in the uniformly distributed case. As a consequence, a high spatial diversity value for a tract suggests that people tweeting from that tract split their tweets evenly among all the tracts they are tweeting from. In contrast, a low value implies that people tweeting from that tract concentrate their tweets in few tracts.

\section{Results}
\label{sec:results}
Happier places tend to attract more Twitter users: the correlation between the mean valence of tweets posted from a tract and the number of people tweeting from the tract is 0.30 ($p<0.001$). Beyond this correlation, we observe systematic trends between emotions expressed in tweets posted from a tract, and the structure of social interactions of people tweeting from that tract.


A number of variables seem to contribute to shaping the structure of online social interactions within an urban area.
Amongst these possibilities are \emph{affect} and \emph{demographic} factors.  The \emph{affect} variables  are: valence ($V$), arousal ($A$), dominance ($D$), positive ($P$), negative ($N$) sentiment. The \emph{demographic} variables are: mean income ($inc$) of population within a tract, education ($edu$) as measured by the percentage of population with bachelor's degree of above, percentage of employed residents ($emp$), their median age ($age$), and the fraction of Hispanic ($hsp$), Asian ($asi$), African American ($blk$), populations white ($wht$) populations within a tract.
The correlations of these variables with tie strength ($tie$) and spatial diversity ($spa$) are described in Tables \ref{tbl-strength-of-ties-affect} and \ref{tbl-strength-of-ties-census}.

Although not presented in these Tables, another interesting relation is between valence and dominance: they are highly and significantly correlated ($r=0.88$, $p<0.001$).
Bearing these relations in mind, we test linear regression models for tie strength ($tie \sim V \times D+A$, $tie \sim P+N$, $tie \sim inc+edu+age+emp$ and $tie \sim hsp+wht+blk+asi$) and spatial diversity ($spa \sim V \times D+A$, $spa \sim P+N$, $spa \sim inc+edu+age+emp$ and $spa \sim hsp+wht+blk+asi$).

\begin{figure*}[t!]
  \centering
  \includegraphics[width=2.0\columnwidth]{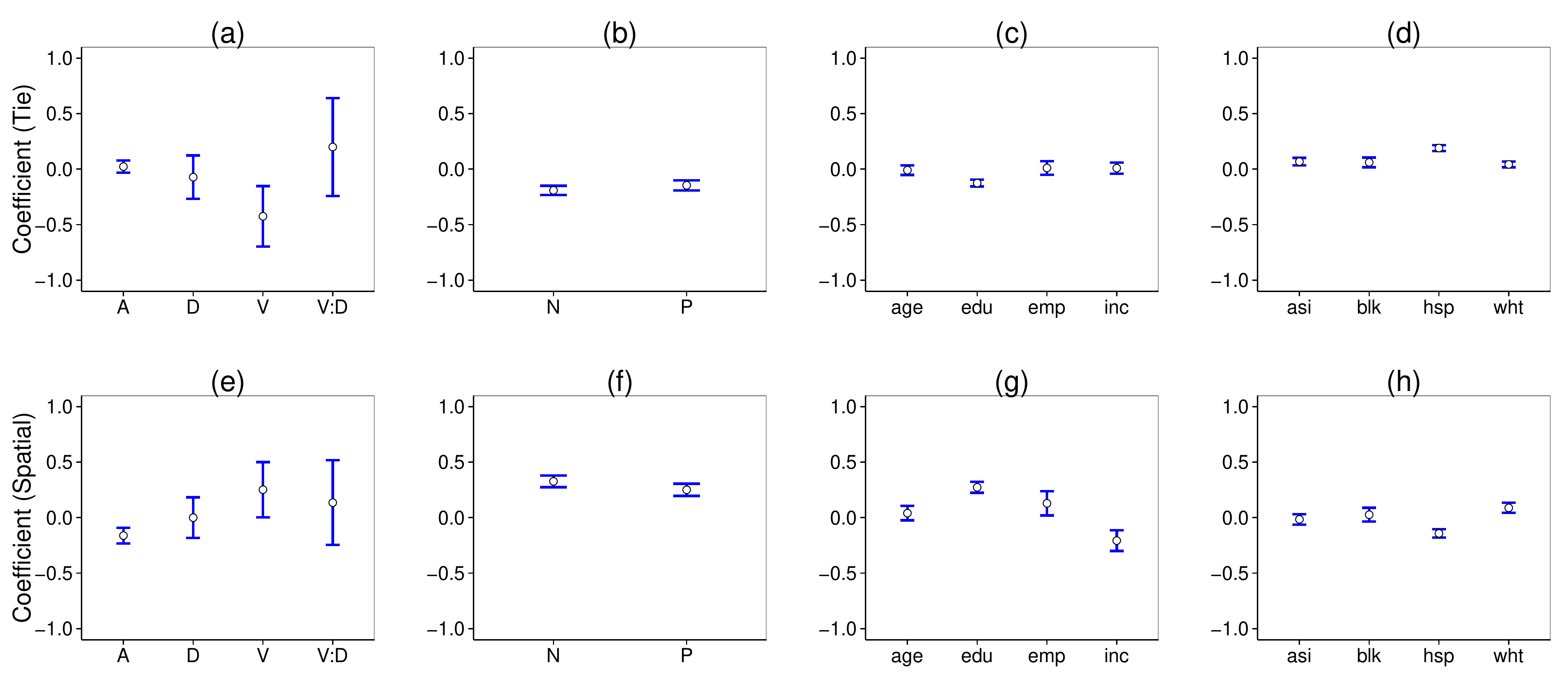}
  \caption{Error bars based on linear regression models: coefficients (tie, spa) are estimated against sentiment (WBK (a), SentiStrength (b)), socioeconomics (c) and ethnicity (d) variables}\label{fig:lm_plot}
\end{figure*}

Regression coefficients for each model are computed on normalized data and summarized in Figure~\ref{fig:lm_plot}.
For tie strength (top row), some interesting results are depicted.
In (a), while arousal and valence:dominance are mainly neutral, dominance and valence have negative dependencies with strength of ties: i.e., the lower the valence, the stronger the tie strength.
Similarly, in (b), negative and positive sentiment represent negative dependencies with tie strength. Hence, larger positive sentiment ($P$) values are associated with weaker social ties.
Stronger ties are associated with smaller values of $N$, i.e., more negative sentiment.
In (c), education has a negative dependence, while others are basically neutral and, in (d), Hispanic ethnicity has a positive dependence to tie strength, with others basically neutral.

Regarding spatial diversity (bottom row in Fig.~\ref{fig:lm_plot}), results are more diverse.
In (e), dominance and valence:dominance are mainly neutral, whereas arousal presents negative dependence and valence presents positive dependence, i.e. the higher the valence, the broader the spatial diversity.
In (f), negative and positive sentiment have positive dependencies with respect to spatial diversity.
In (g), age is neutral, education and employment percentage present positive and mean income negative dependencies.
Finally, in (h), besides Hispanic presenting negative dependence, all others are mainly neutral.

Overall, these results show that besides the correlations between the chosen variables, some of them contribute more to explain the tie strength and social diversity while others have minor contributions.
More details regarding affect, demographics and social interactions results are presented in the following.

\subsection{Emotion and Social Ties}
Table~\ref{tbl-strength-of-ties-affect} reports correlations between affect and the structure of social interactions measured from tweets. We quantify affect using mean valence, arousal and dominance of tweets posted from a tract (measured by WKB lexicon), and mean positive and negative sentiment, (measured by SentiStrength). The social variables are average social tie strength per tract and spatial diversity, a measure of inter-tract mobility. Most of these correlations are statistically significant.

\begin{table}[tbph]
  \centering
  \begin{tabular}{lcc}
    \toprule
     \textbf{Variables} & \textbf{Tie Strength}  & \textbf{Spatial Diversity} \\ 
     \midrule
    Valence   & -0.36$^{***}$ & 0.32$^{***}$ \\
    Arousal   & 0.14$^{***}$  & -0.20$^{***}$ \\
    Dominance & -0.31$^{***}$ & 0.31$^{***}$ \\ 
    \midrule
    Positive Sent. & -0.18$^{***}$ & 0.23$^{***}$ \\
    Negative Sent. & -0.24$^{***}$ & 0.28$^{***}$ \\
    \bottomrule
    \multicolumn{3}{l}{\textit{\small{*p<0.05, **p<0.01, ***p<0.001}}}\\ 
  \end{tabular}
  \caption{Correlation coefficient of per-tract strength of ties measure and mean value of affect. }\label{tbl-strength-of-ties-affect}
\end{table}

\begin{figure}[tbph]
  \centering
  \includegraphics[width=0.8\columnwidth]{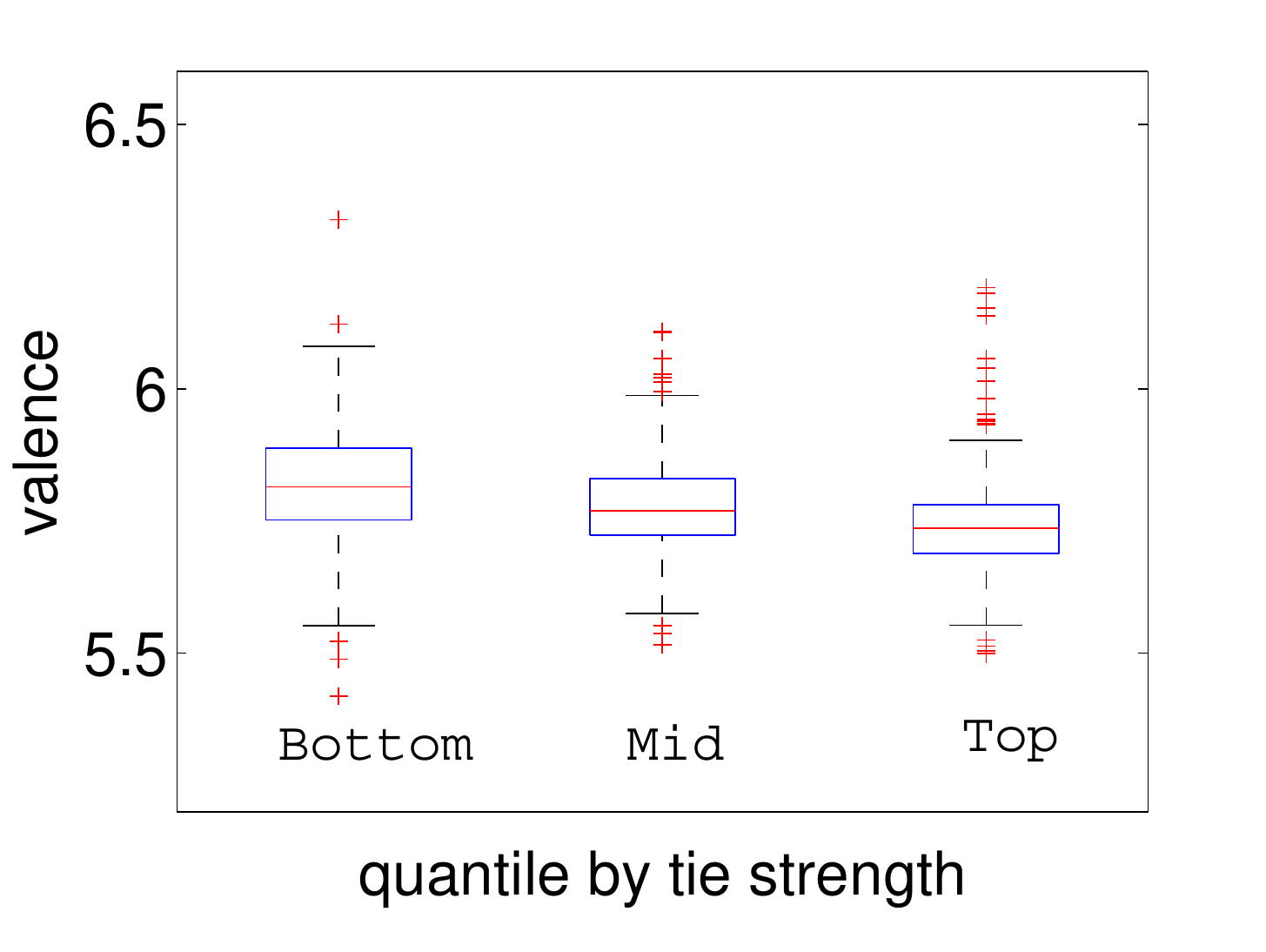}
  \caption{Box plot of mean valence of tweets posted from tracts, after they were grouped by mean tie strength into three equal-sized groups, or tertiles. Lines inside the boxes indicate median value of valence, whereas the boxes themselves show interquartile ranges. }\label{fig:valence-tie-boxplot}
\end{figure}


The average value of valence across all tracts is 5.78, which corresponds to slightly positive affect
with respect to the neutral point of 5.0, in line with emotional expression in other media~\cite{Garcia2012b}. This contrasts with the negative correlation between valence and tie strength. Consequently, tracts with stronger social ties are associated with less positive --- sadder, angrier --- tweets than tracts with weaker social ties, which are associated with more positive --- happier --- tweets. This can be seen clearly in Figure~\ref{fig:valence-tie-boxplot}, which shows the values of mean valence for tracts after they were ordered by average social tie strength and divided into three equal-sized bins, or tertiles. The bottom third of tracts, i.e., those with weakest social ties, have the highest values of valence (mean 5.82). In contrast, the top tertile composed of tracts with strongest social ties has the lowest values of valence (mean 5.74). These differences are statistically significant ($p<0.01$).

\begin{figure*}[t!]
  \centering
  \begin{tabular}{ccc}  
  \includegraphics[width=0.65\columnwidth]{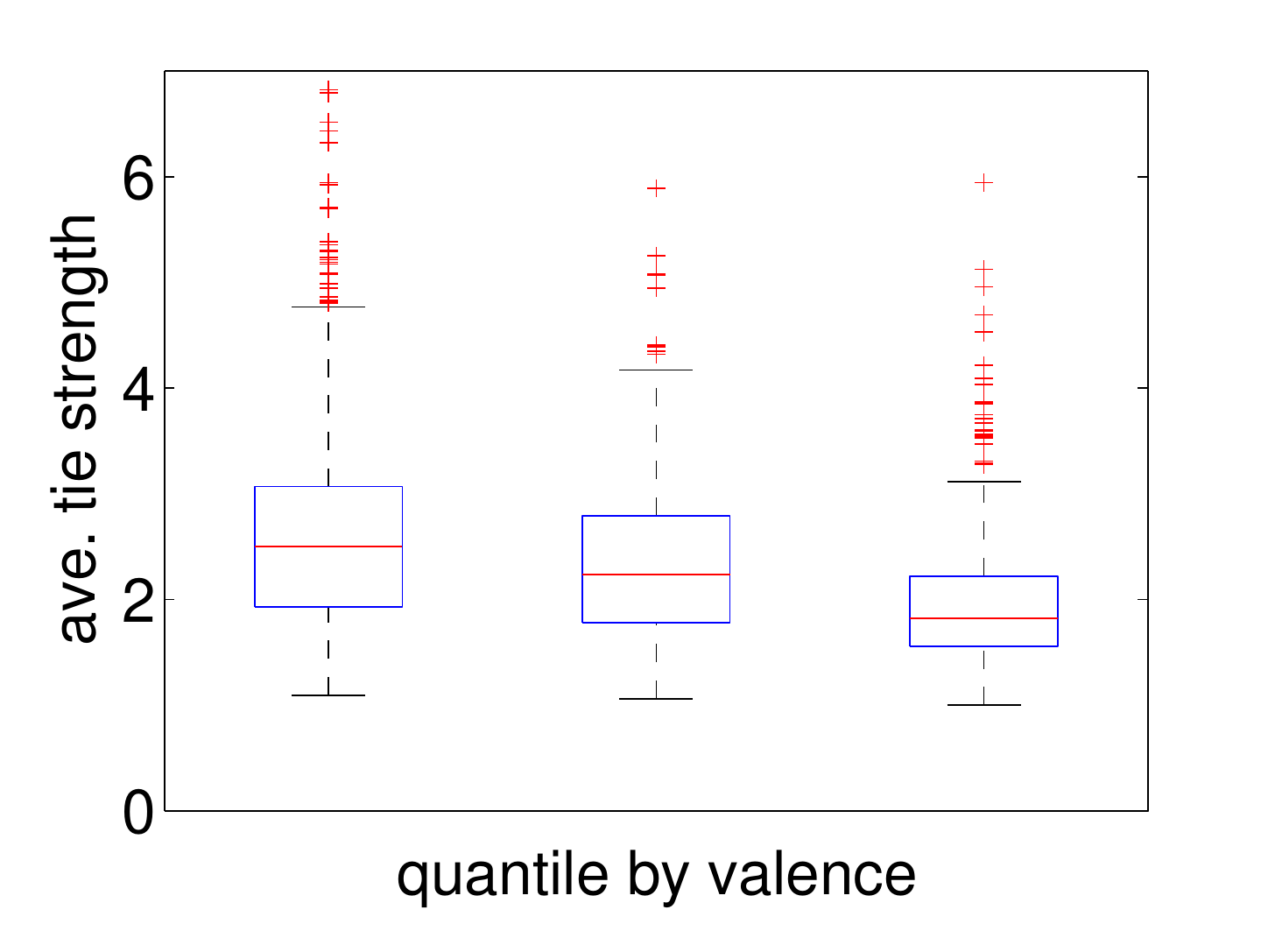} &
  \includegraphics[width=0.65\columnwidth]{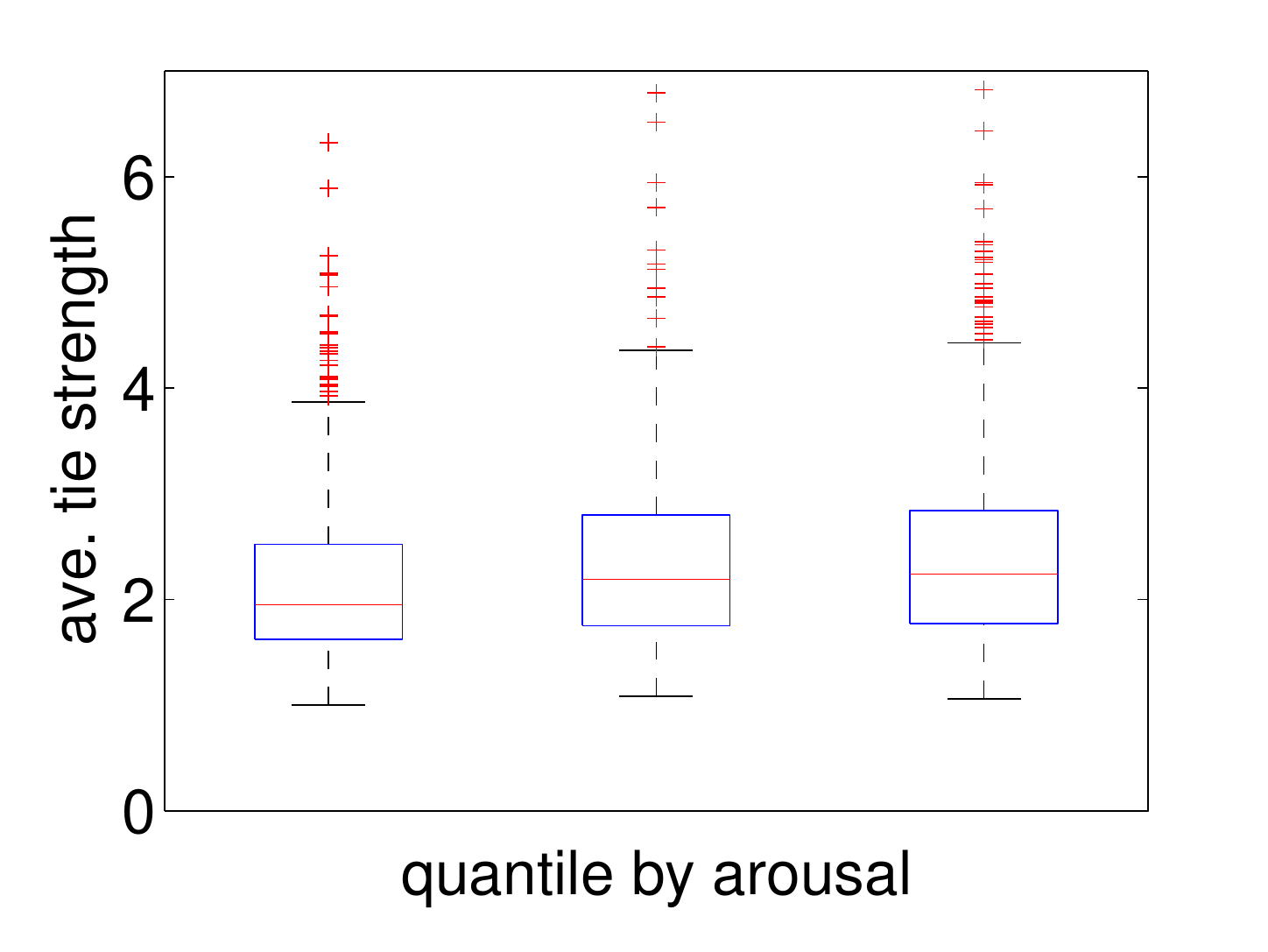} &
  \includegraphics[width=0.65\columnwidth]{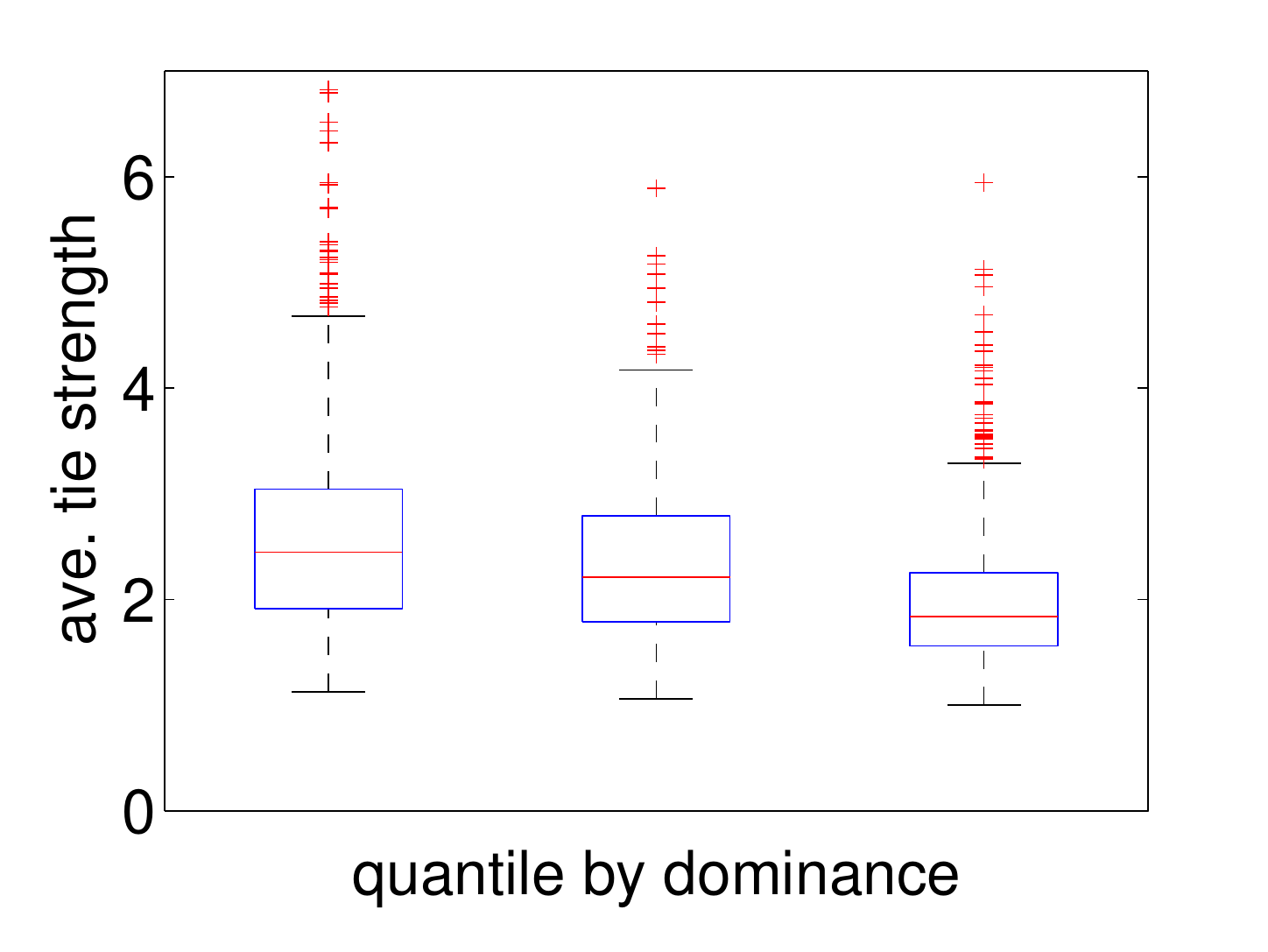}
  \end{tabular}
  \caption{Social tie strength analysis. Tracts were ordered by the mean value of (a) valence, (b) arousal, or (c) dominance of tweets from that tract and divided into three groups of equal size.  Lines inside the boxes indicate median value of tie strength for each group, whereas the boxes show interquartile ranges. }\label{fig:tie-strength-quintiles}
\end{figure*}

For a different perspective on this relationship, we order tracts by the average valence, arousal, and dominance of their tweets and group them into three equal-size bins. Figure~\ref{fig:tie-strength-quintiles} reports the average tract tie strength across these tertiles.
Tracts from which people post lower valence (sadder) tweets, on average, are associated with significantly stronger social ties ($p<0.01$) than tracts with higher valence (happier) tweets (Figure~\ref{fig:tie-strength-quintiles}(a)). Tracts from which users post messages expressing higher arousal tend to have slightly stronger ties (Figure~\ref{fig:tie-strength-quintiles}(b)) although only the difference between mean tie strength of the mid and top tertiles are significant ($p<0.05$). In contrast, tracts associated with message expressing higher dominance, Figure~\ref{fig:tie-strength-quintiles}(b), are associated with significantly weaker social ties ($p<0.01$).
Since weak ties are associated with more diverse social relationships, our observation confirms the relationship between emotions and diversity of network structure.

The sentiment values computed by SentiStrength are consistent with this trend. The correlation between tie strength and positive sentiment while weaker, is also negative, suggesting that tracts with more positive tweets have weaker ties. Note that values for negative sentiment are below zero, with lower values representing stronger negative sentiment. In this case, negative correlation with tie strength means that tracts with more negative tweets have stronger ties than tracts with less negative tweets.
This result is in line with theories of social regulation of emotions~\cite{Rime2009} and with previous results in protest movements that showed how online negative emotions were associated with stronger collective action~\cite{Alvarez2015}.

%
%

\begin{figure}[tbph]
  \centering
  \includegraphics[width=0.7\columnwidth]{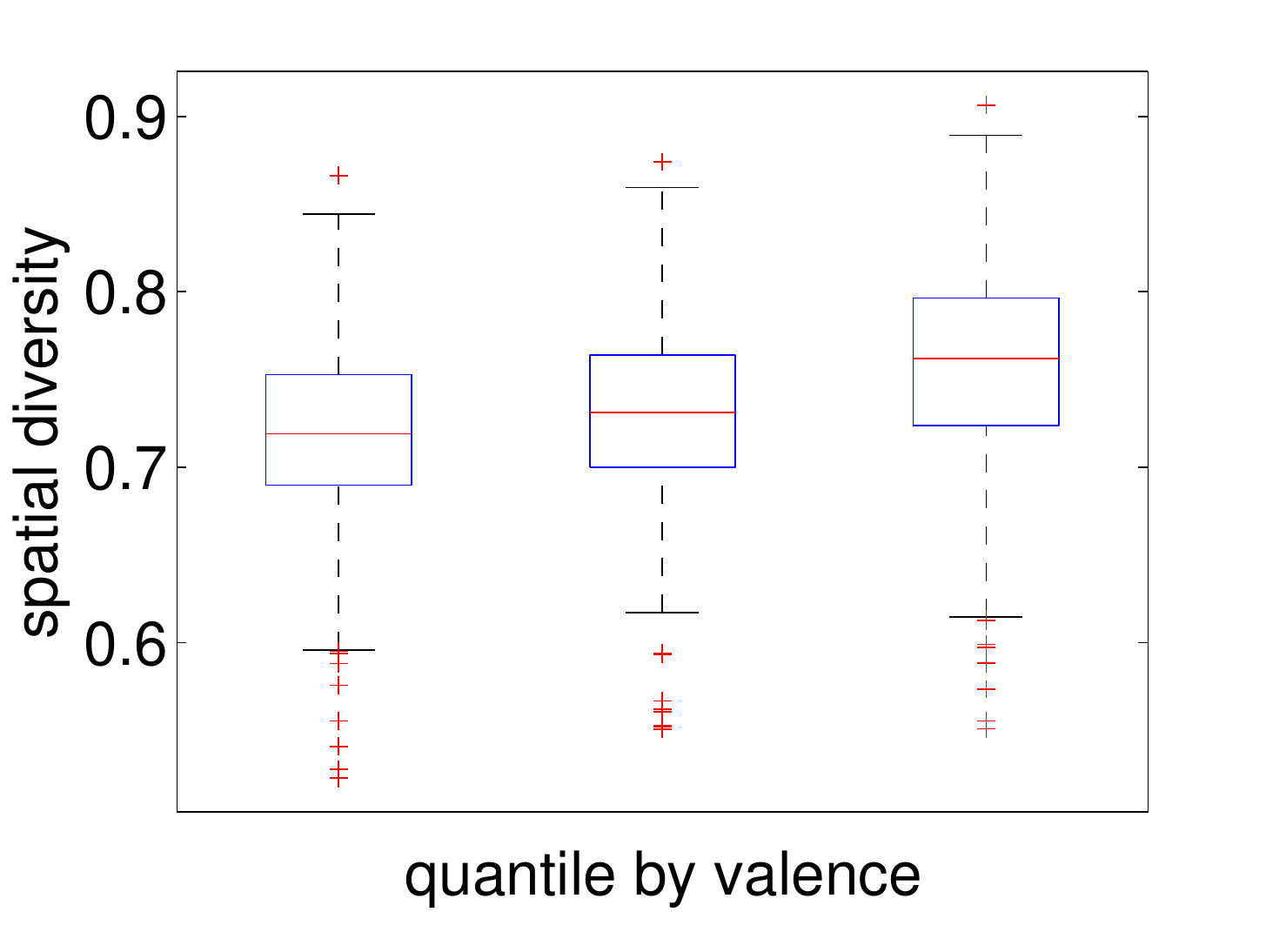}
  \caption{Spacial (inter-tract) diversity for three quantiles, ordered by the mean valence of tweets posted from them. }\label{fig:mobility-valence}
\end{figure}


We also report the relationship between affect of a tract and the spatial diversity of users tweeting from that tract, a measure of inter-tract mobility defined in Section~\ref{sec:method-sna}. As shown in Figure~\ref{fig:mobility-valence}, tracts with more positive tweets have significantly higher spatial diversity than tracts with less positive tweets ($p<0.001$).
This suggests that people expressing happier emotions move (and tweet from) a larger number of different places than people expressing more negative emotions, whose movements are confined to a smaller set of tracts.

\begin{figure*}[tbph]
  \centering
  \begin{tabular}{ccc}  
  \includegraphics[width=0.65\columnwidth]{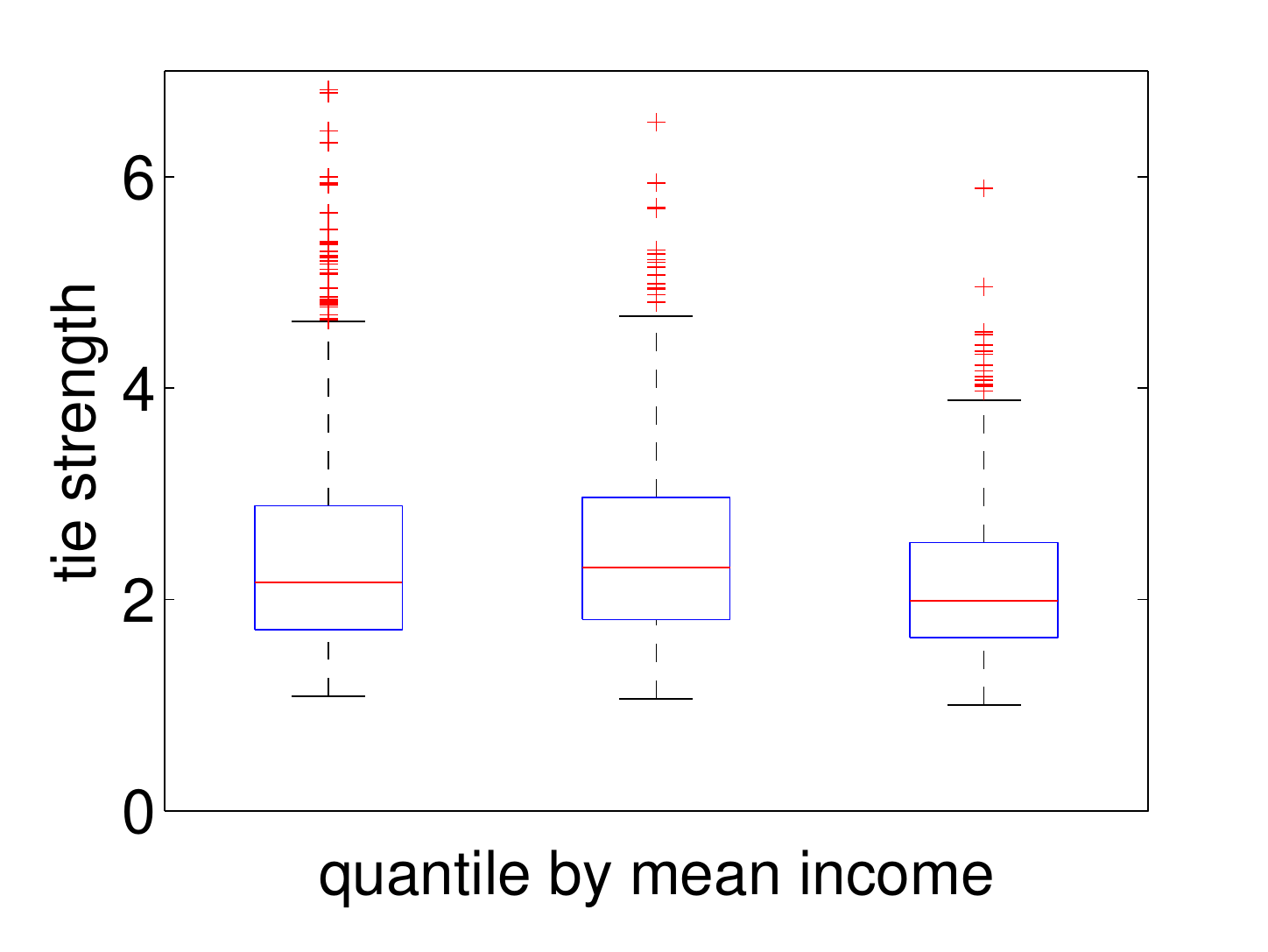} &
  \includegraphics[width=0.65\columnwidth]{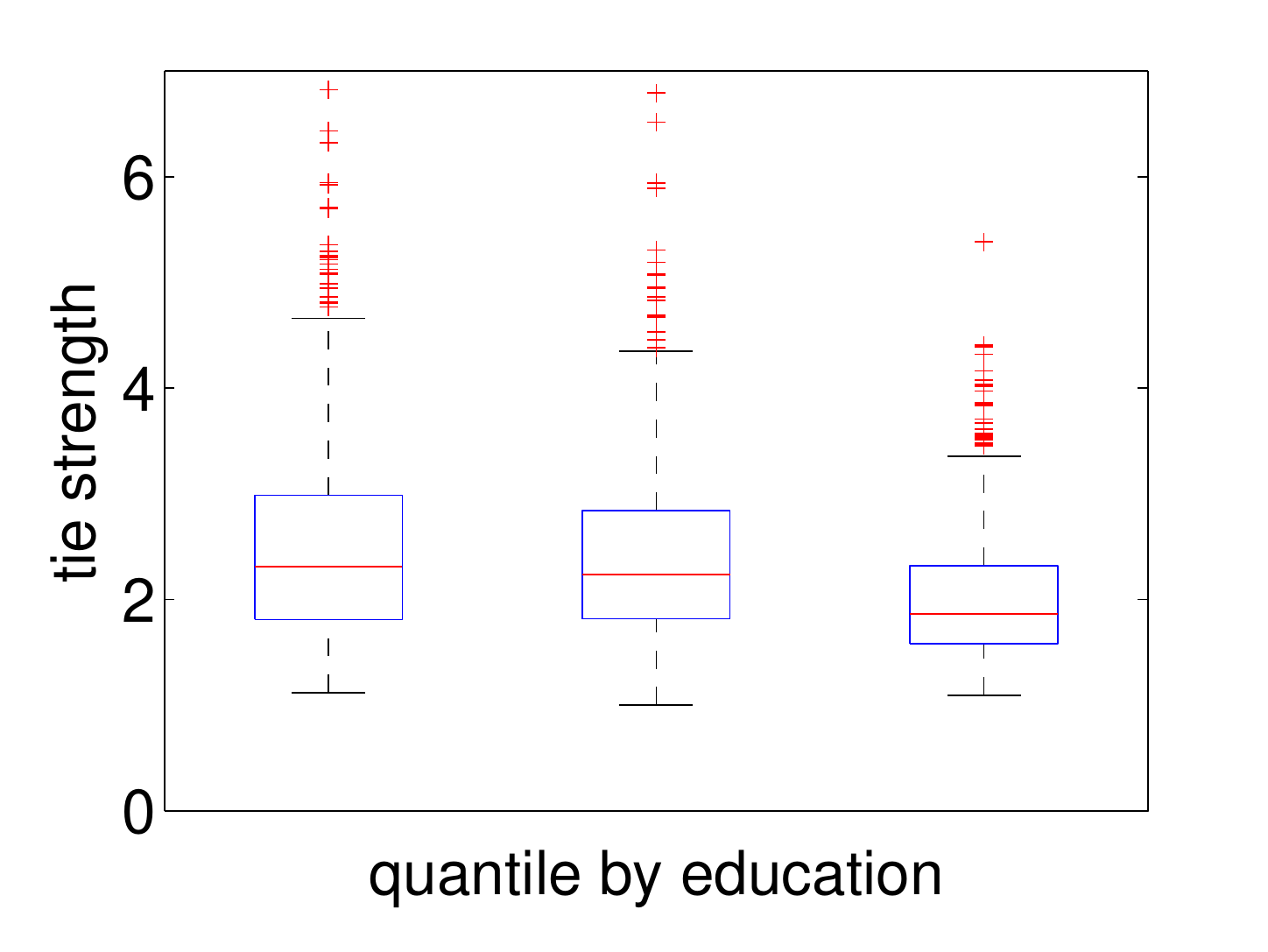}&
  \includegraphics[width=0.65\columnwidth]{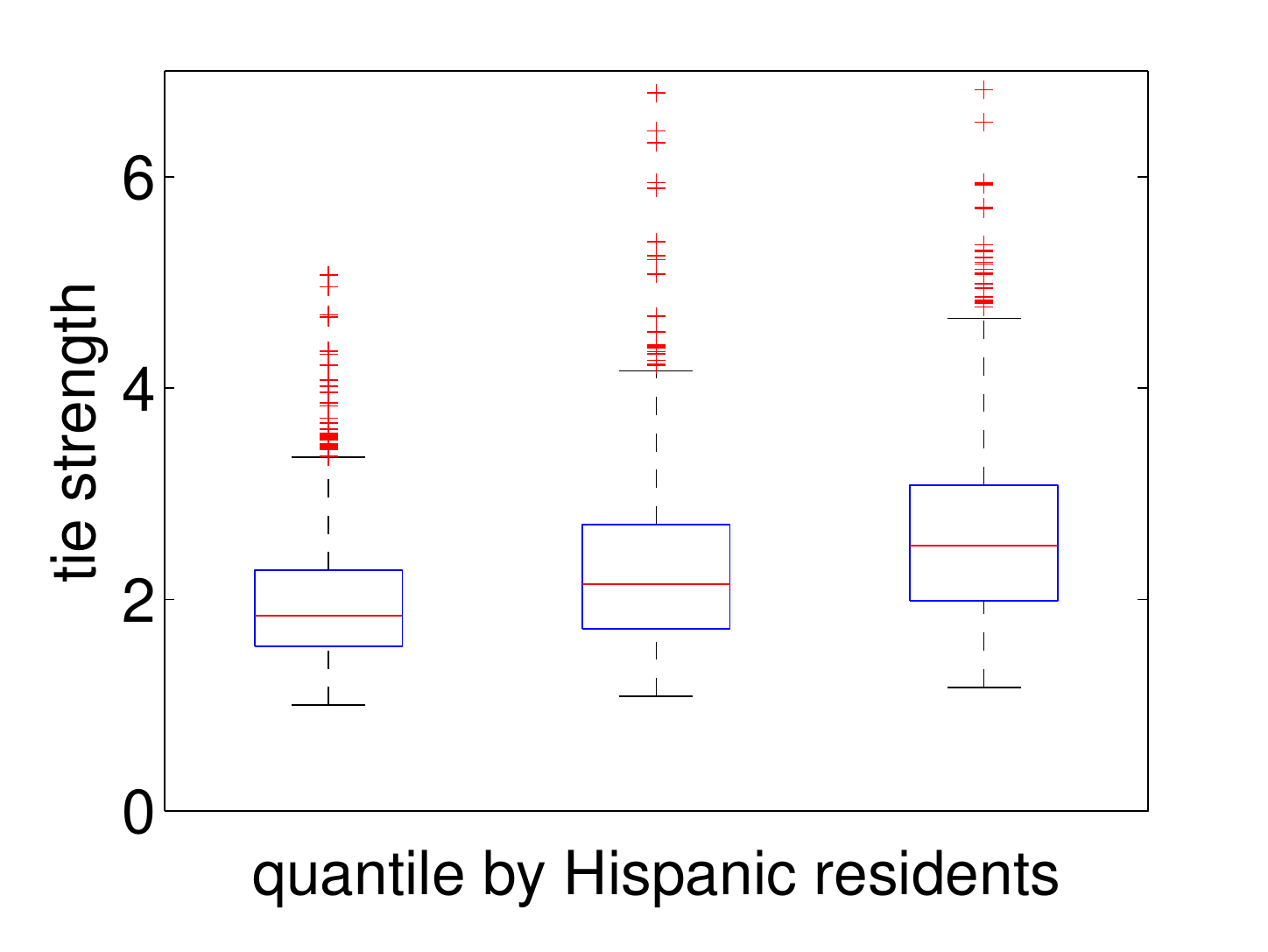}
  \end{tabular}
  \caption{Demographic analysis of social ties. Tracts were ordered by demographic characteristic (mean household income, percentage of residents with Bachelor's degree or above, and number of Hispanic residents) in that tract and divided into three groups of equal size. Lines inside the boxes indicate median value of tie strength within each group, whereas the boxes show interquartile ranges. Leftmost box represents the bottom third of tracts with lowest values of the demographic attribute, and rightmost box represents the top third of tracts with largest values of the attribute. }\label{fig:tie-strength-demo}
\end{figure*}

\subsection{Demographics and Social Ties}
We find a relationship between demographic characteristics of tracts from which people tweet and the social structure of tweeting users. Table~\ref{tbl-strength-of-ties-census} reports correlations between average per-tract social tie strength, spatial diversity and demographic variables extracted from US Census data. While a previous study~\cite{Eagle:Network2010} of the relationship between social structure and economic prosperity used somewhat different indicators --- e.g., they used network diversity to measure diversity of social contacts --- we reach qualitatively similar conclusions. Specifically, we find a negative correlation between (mean) income, one measure of economic prosperity, and sociability: higher incomes are associated with weaker social ties. There is an even stronger negative relationship between education (measured by the percentage of tract residents who hold a Bachelor's degree or higher) and sociability. Socioeconomic variables are highly inter-correlated, and it is difficult to infer causal relationships within data. Still, the higher correlation of education with tie strength and mobility implies that education may be more predictive of sociability than other demographic attributes.

\begin{table}
  \centering
\begin{tabular}{lcc}
  \hline
  \textbf{Variables} & \textbf{Tie Strength} & \textbf{Spatial Diversity} \\ 
  \toprule
  Mean Income & -0.12$^{***}$  & 0.08$^{**}$ \\
  Employment & -0.12$^{***}$   & 0.23$^{***}$ \\
  Age        & -0.18$^{***}$  & 0.18$^{***}$ \\
  Education  & -0.27$^{***}$  & 0.35$^{***}$ \\
  \midrule
  White    & -0.14$^{***}$ & 0.20$^{***}$  \\
  African Amer.    & 0.06$^{*}$   & 0.03          \\
  Hispanic & 0.35$^{***}$  & -0.27$^{***}$ \\
  Asian Amer.   & -0.03         & 0.10$^{***}$    \\
  \bottomrule
  \multicolumn{2}{l}{\small{*p<0.05, **p<0.01, ***p<0.001}}
\end{tabular}  \caption{Correlation between demographic characteristics of tracts and measures of social structure of people tweeting from these tracts.  }\label{tbl-strength-of-ties-census}
\end{table}

We take a closer look at the relationships between social tie strength and demographic attributes using quantile analysis. Figure~\ref{fig:tie-strength-demo} shows the box plot of tie strength of a group of tracts after they were divided into three equal-size groups based on each demographic attribute. For example, Figure~\ref{fig:tie-strength-demo}(top) divides tracts by mean income into the lowest-earning, mid-earning, and highest-earning tertiles and shows the range of social tie strength values in each tertile. Only the top income and education tertiles have significantly weaker social ties than the other two tertiles ($p<0.01$). However, when tracts are divided into tertiles according to the number of Hispanic residents, the differences between the mean tie strengths of all tertiles are significant ($p<0.01$). This result highlights the social component of culture: Hispanic cultures
focus more on sociability values and are  less individualist than anglo-saxon cultures
(for example, Mexico scores 30 and the US 91 in the individualism scale of Hofstede~\cite{Hofstede1980}). This provides an explanation for the stronger links of tracts with higher number of Hispanic residents, as their online network structures reflect their shared values~\cite{Garcia2013cultural,Kayes2015}.

\begin{figure*}[tbph]
  \centering
  \begin{tabular}{ccc}  
  \includegraphics[width=0.65\columnwidth]{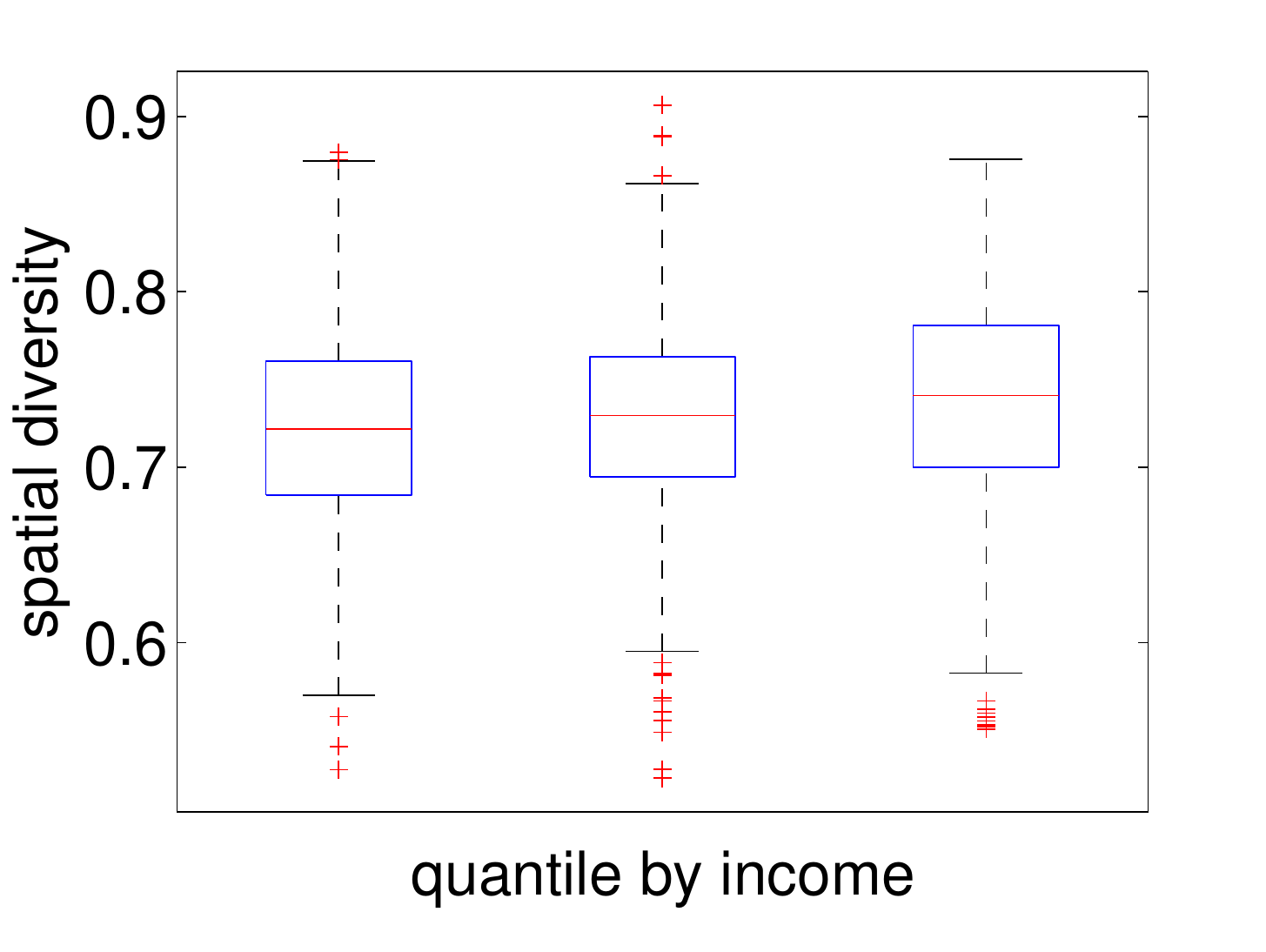} &
  \includegraphics[width=0.65\columnwidth]{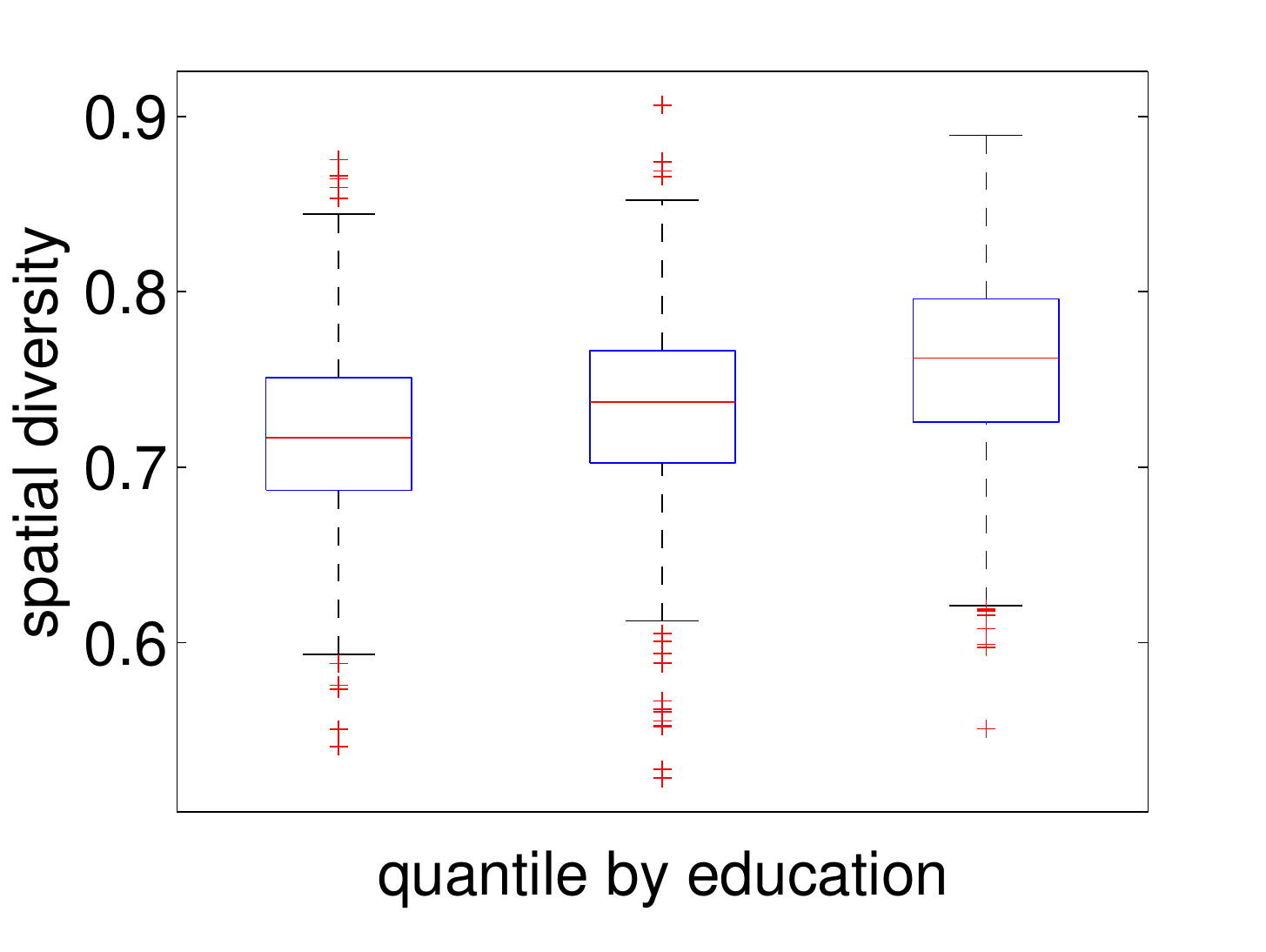}&
  \includegraphics[width=0.65\columnwidth]{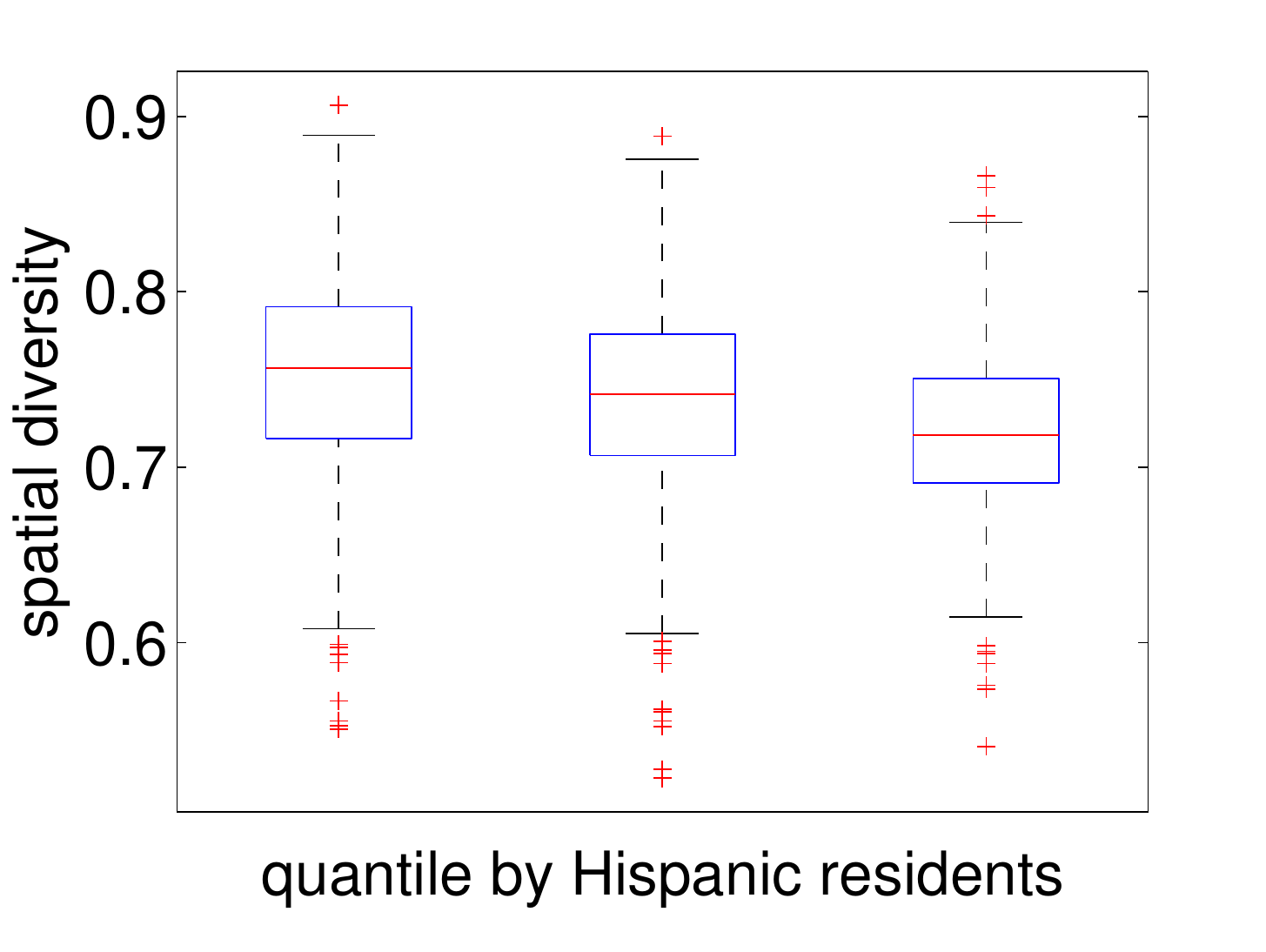}
  \end{tabular}
  \caption{Demographic analysis of spatial diversity (inter-tract mobility). Tracts were ordered by (a) mean household income, (b) percentage of residents with Bachelor's degree or above, or (c) number of Hispanic residents in that tract and divided into three groups of equal size. The figure shows the box plot of inter-tract mobility values of tracts in different quantiles. Leftmost box represents the bottom third of tracts with lower values of the demographic attribute, and rightmost box represents the top third of tracts with largest values of the attribute. }\label{fig:mobility-demo}
\end{figure*}

Figure~\ref{fig:mobility-demo} shows the association between spatial diversity and demographic characteristics. Income does not appear to significantly affect spatial diversity: only the top tertile of tracts by incomes has a significantly different spatial diversity ($p<0.001$) from the other two tertiles. Education, however, has a stronger dependence:  tracts with better-educated residents also have significantly higher ($p<0.001$) spatial diversity than tracts with fewer educated residents. In addition, ethnicity appears to be a factor. Tracts with larger Hispanic population have significantly lower spatial diversity ($p<0.01$) than other tracts.


\section{Discussions} 
\label{sec:discussion}
The availability of large scale, near real-time data from social media sites such as Twitter brings novel opportunities for studying online behavior and social interactions at an unprecedented spatial and temporal resolution. By combining Twitter data with US Census, we were able to study how the socioeconomic and demographic characteristics of residents of different census tracts are related to the structure of online interactions of users tweeting from these tracts. Moreover, sentiment analysis of tweets originating from a tract revealed a link between emotions and sociability of Twitter users.

Our findings are broadly consistent with results of previous studies carried out in an offline setting, and also give new insights into the structure of online social interactions. We find that at an aggregate level, areas with better educated, somewhat younger and higher-earning population are associated with weaker social ties and greater spatial diversity (or inter-tract mobility). In addition, Twitter users express happier, more positive emotions from these areas. Conversely, areas that have more Hispanic residents are associated with stronger social ties and lower spatial diversity. People also express less positive, sadder emotions in these areas. Since weak ties are believed to play an important role in delivering strategic, novel information, our work identifies a social inequity, wherein the already privileged ones (more affluent, better educated, happier) are in network positions that potentially allow them greater access to novel information.

Some important considerations limit the interpretation of our findings. First, our methodology for identifying social interactions may not give a complete view of the social network of Twitter users. Our observations were limited to social interactions initiated by users who geo-reference their tweets. This may not be representative of all Twitter users posting messages from a given tract, if systematic biases exist in what type of people elect to geo-reference their tweets.
For demographic analysis, we did not resolve the home location of Twitter users. Instead, we assumed that characteristics of an area, i.e., of residents of a tract, influence the tweets posted from that tract.
Other subtle selection biases could have affected our data and the conclusions we drew~\cite{Tufekci2014}. It is conceivable that Twitter users residing in more affluent areas are less likely to use the geo-referencing feature, making our sample of Twitter users different from the population of LA county residents. Recognizing this limitation, we did not make any claims about the behavior of LA residents; rather, we focused on 
the associations between emotions and characteristics of a place and the behavior of Twitter users, with an important caveat that those who turn on geo-referencing may differ from the general population of Twitter users.

For the analysis of emotions, we only considered English language tweets, although a significant fraction of tweets were in Spanish. This may bias the average affect of tracts, especially for low-valence tracts, which have a larger number of Hispanic residents. In the future, we plan to address this question by conducting sentiment analysis of Spanish language tweets.

\section{Acknowledgments}
MA was supported by the USC Viterbi-India internship program. LG acknowledge support by the National Counsel of Technological and Scientific Development --- CNPq, Brazil (201224/2014--3), and USC-ISI visiting researcher fellowship. This work was also partially supported by DARPA, under contract W911NF-12-1-0034.
This support is gratefully acknowledged.


\end{document}